\documentclass[3p,times]{elsarticle}

\usepackage{ecrc}
\usepackage{subfigure}

\volume{00}

\firstpage{1}

\journalname{Nuclear Physics A}

\runauth{}

\jid{nupha}

\usepackage{amssymb}

\usepackage[figuresright]{rotating}
\usepackage[rightcaption]{sidecap}

\begin{document}

\begin{frontmatter}



\dochead{}

\title{Azimuthal angular correlations between heavy-flavour decay electrons and charged hadrons in pp collisions at $\sqrt{s} = 2.76$ TeV in ALICE}


\author{Deepa Thomas on behalf of the ALICE collaboration}

\address{ERC- Research Group QGP-ALICE, Utrecht University, Princetonplein 5,\\ 3584 CC Utrecht, The Netherlands}
\ead{Deepa.Thomas@cern.ch}
\begin{abstract}

We present the measurement of azimuthal angular correlations between electrons and charged hadrons in pp collisions at 2.76 TeV measured with the ALICE experiment at the LHC. The correlation distributions from PYTHIA simulations are used to extract the relative contribution from B-hadron decays to the yield of electrons from heavy-flavour decays up to $p_\textrm{t}$ = 10 GeV/c.

\end{abstract}

\begin{keyword}

Heavy-flavour production, Azimuthal angular correlations, Large Hadron Collider
\end{keyword}

\end{frontmatter}

\label{introduction}
Heavy quarks, produced in the initial hard scattering processes, probe the high energy density Quantum Chromodynamic (QCD) matter that is formed in the high-energy heavy ion collisions. Heavy-flavour hadron production can be studied by measuring electrons produced through semi-leptonic decays. To understand the energy loss of heavy quarks in the QCD medium, it is essential to disentangle the electrons from decay of D and B mesons. In pp collisions the measurement of the contribution from beauty to the heavy-flavour decay electron spectrum provides a baseline for measurements in Pb-Pb collisions, and a test for perturbative QCD calculations of heavy quark production.

The shape of the azimuthal angular correlations of heavy-flavour decay electrons and charged hadrons can be used to determine the relative beauty contribution. The width of the near side correlation distribution is larger for B mesons compared to D mesons due to different decay kinematics.

\begin{SCfigure}[1.2][h]
\centering
\includegraphics[height=5.4cm,width=6.0cm]{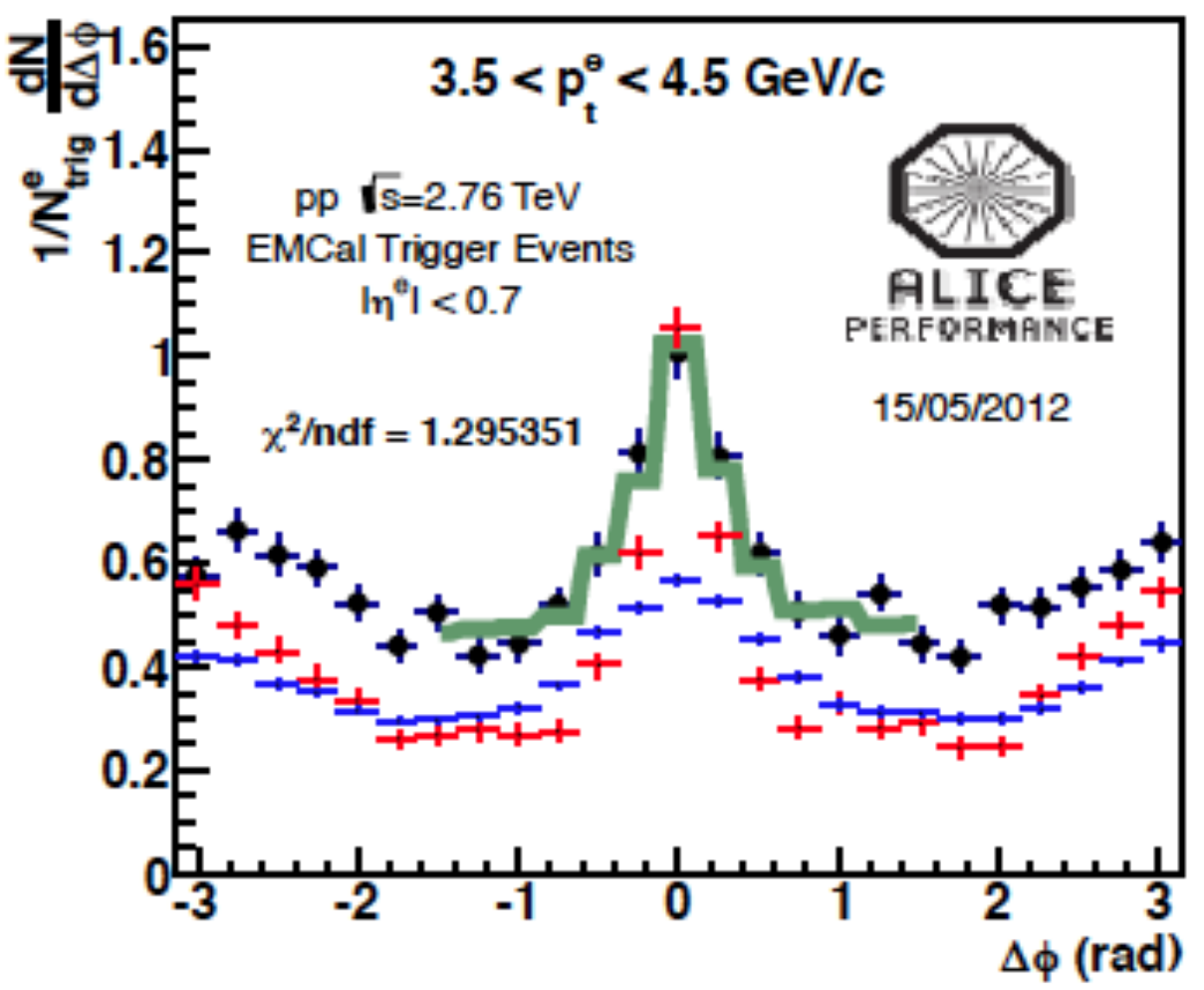}
\label{fig:deltaphifit}
\caption{Azimuthal angular correlation between heavy-flavour decay electrons and charged hadrons in pp collisions. The red histogram is the MC distribution for electrons from charm decay, blue histogram is the MC distribution for electrons from beauty decay. The full green curve is the fit to the data (black) points.}
\end{SCfigure}

\begin{figure}[h]                                                                        
\centering
\subfigure{
    \includegraphics[height=6cm,width=8cm]{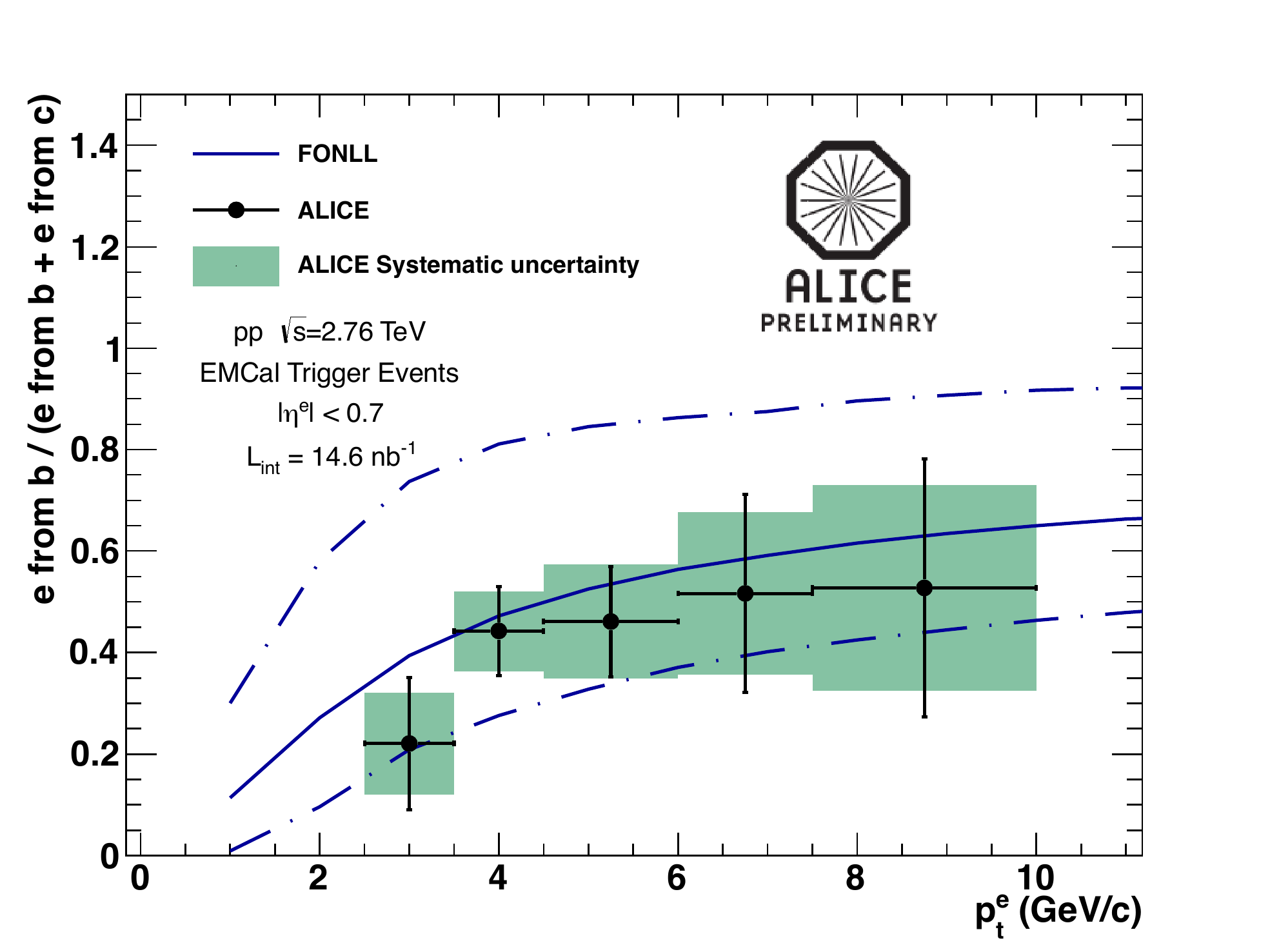}                                                
}
\subfigure{                                                                               
    \includegraphics[height=6cm,width=8cm]{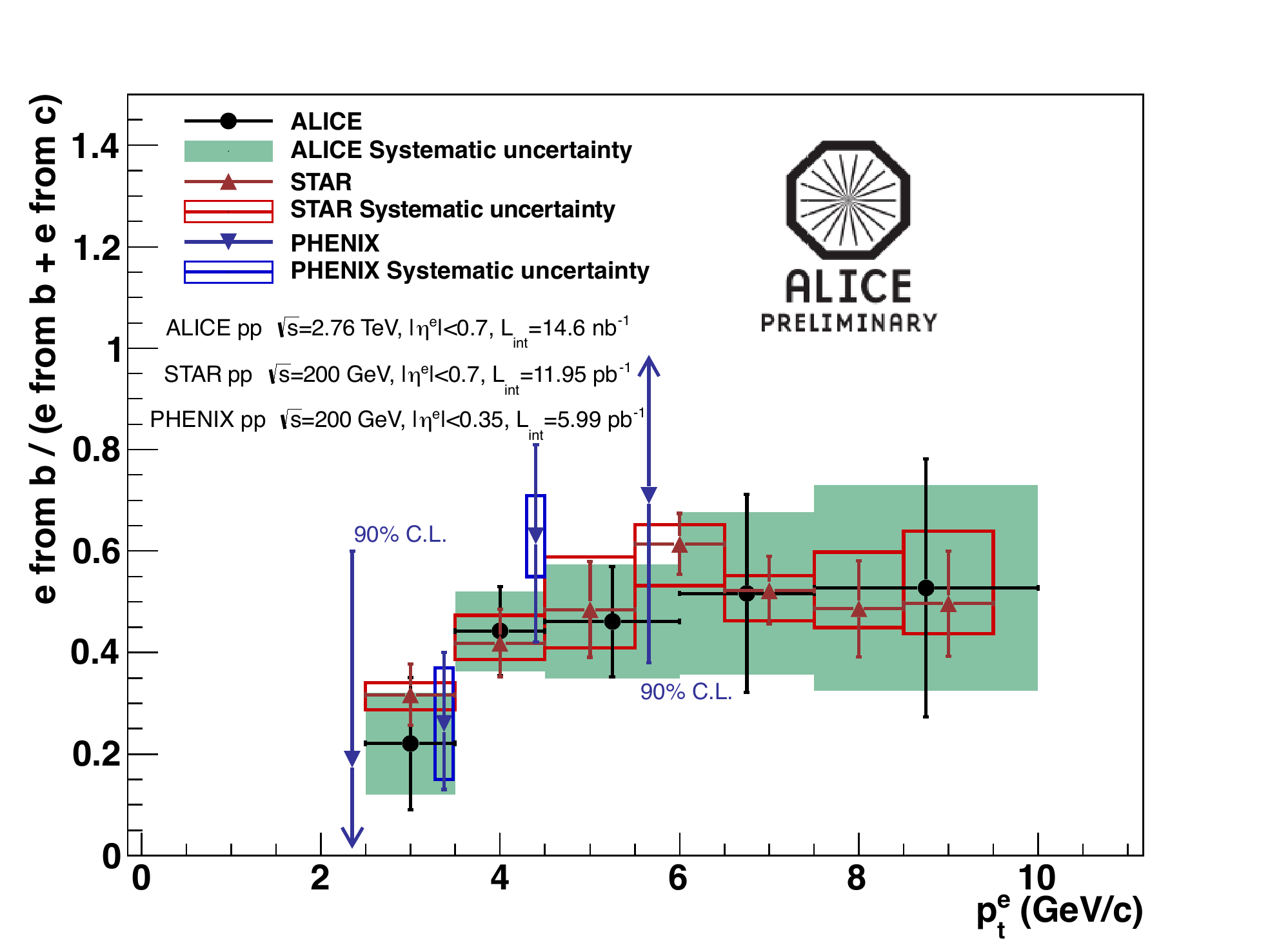}                                           
} 
\caption[]{Relative beauty contribution to the heavy-flavour decay electron yield compared with FONLL calculation and RHIC measurements }
\label{fig:bratio}                                                              
\end{figure}

The analysis is performed for pp collisions at 2.76 TeV centre-of-mass energy collected in March 2011 with the ALICE experiment ~\cite{aliceDet}. For the analysis, the detectors used are the Inner Tracking System (ITS) ($|\eta|<0.9$, $0<\phi<360^{\circ}$), the Time Projection Chamber (TPC) ($|\eta|<0.9$, $0<\phi<360^{\circ}$) and the Electromagnetic Calorimeter (EMCal) ($|\eta|<0.7$, $80<\phi<180^{\circ}$). The events which pass EMCal L0 trigger conditions are studied. The L0 trigger is a 2$\times$2 tower patch with a trigger cluster energy threshold of 3 GeV.

Electrons are identified using information from TPC and EMCal detectors of the ALICE experiment. Particle identification in the TPC is based on the measurement of specific ionization energy loss in the detector gas. In the EMCal, electron candidates are required to have $E/p$ between 0.8 and 1.2. The non-heavy flavour electrons (Non-HFE) are identified using the invariant mass method. For the invariant mass calculation, the partner electron is selected after loose electron identification cuts. Electron pairs which have an invariant mass of less than 50 MeV/$\textrm{c}^{2}$ are tagged as Non-HFE. Above an electron $p_{\textrm{t}}$ of 2 GeV/c, the non-HFE finding efficiency is $\approx 50\%$, estimated from Monte Carlo simulations. The remaining Non-HFE contamination in the HFE sample is corrected using the efficiency.

The azimuthal angular correlations between heavy-flavour decay electron and charged hadrons is constructed with heavy-flavour electrons and tracks which pass some quality checks. To determine the ratio of electrons from beauty decay, the measured correlation distribution is fit with the function

\begin{equation}
\Delta\phi_{e-h}^{HF} = const + r_{B} \Delta\phi_{e-h}^{B} + (1-r_{B}) \Delta\phi_{e-h}^{D},
  \end{equation}

where  $r_{B}$ = $\frac {e_{B}} {e_{B} + e_{D}}$ is the ratio of electron yield from B-meson decays to that of the heavy-flavour electron yield and $const$ gives the uncorrelated background. $\Delta\phi_{e-h}^{D}$ ($\Delta\phi_{e-h}^{B}$ ) is the azimuthal angular correlation between electron from D (B) meson decay and charged hadrons from Monte Carlo simulations (PYTHIA 6.4 with Perugia-0 tune ~\cite{Pythiatune}) based on a detailed description of the apparatus.

The fitting range used is $-1.5 < \Delta\phi < 1.5$ rad. The correlation distribution and the fit are shown in Figure \ref{fig:deltaphifit}. The beauty ratio extracted from the fit as a function of $p_{\textrm{t}}$ is shown in Figure \ref{fig:bratio}. The $r_{B}$ increases with $p_{\textrm{t}}$ and reaches $\approx 0.5$ ($\frac{e_{B}}{e_{D}} \approx 1$) at around 5 GeV/c. This measurement is in good agreement, within uncertainities, with pQCD FONLL calculations ~\cite{FONLLcurve} and with STAR and PHENIX measurements ~\cite{BottomStar,eh-Bcs-Phenix}. 


\bibliographystyle{elsarticle-num}
\bibliography{<your-bib-database>}

\begin{thebibliography}{00}


\bibitem{aliceDet}
K. Aamodt et al. [ALICE Collaboration], JINST 3, S08002 (2008).
\bibitem{Pythiatune}
P. Skands, arXiv:0905.3418v1 [hep-ph] (2009).
\bibitem{FONLLcurve}
M. Cacciari et al. JHEP 0103, 006 (2001) and private communication (2012).
\bibitem{BottomStar}
M. M. Aggarwal et al. [STAR Collaboration], Phys. Rev. Lett. 105, 202301 (2010).
\bibitem{eh-Bcs-Phenix}
A.Adare et al. [PHENIX Collaboration], Phys. Rev. Lett. 103, 082002 (2009).
\end{thebibliography}

\end{document}